\documentclass[showpacs,pra,aps,superscriptaddress,twocolumn,floatfix]{revtex4}

\usepackage{amsmath}
\usepackage{amssymb,mathrsfs}
\usepackage{graphicx}

\newcommand {\prt}{\partial}
\newcommand{\om}{\omega}

\newcommand{\al}{\alpha}

\newcommand{\ga}{\gamma}

\begin{document}

\title{Wave patterns generated by a flow of two-component Bose-Einstein condensate
with spin-orbit interaction past a localized obstacle}

\author{Y.~V.~Kartashov}
\affiliation{Institute of Spectroscopy,
  Russian Academy of Sciences, Troitsk, Moscow, 142190, Russia}
\author{A.~M.~Kamchatnov}
\affiliation{Institute of Spectroscopy,
  Russian Academy of Sciences, Troitsk, Moscow, 142190, Russia}

\begin{abstract}
It is shown that spin-orbit interaction leads to drastic changes in wave patterns generated by a flow
of two-component Bose-Einstein condensate (BEC) past an obstacle. The combined Rashba and Dresselhaus spin-orbit
interaction affects in different ways two types of excitations---density and polarization waves---which
can propagate in a two-component BEC. We show that the density and polarization ``ship wave'' patterns
rotate in opposite directions around the axis located at the obstacle position and the angle of rotation
depends on the strength of spin-orbit interaction. This rotation is accompanied by narrowing of the
Mach cone. The influence of spin-orbit coupling on density solitons and
polarization breathers is studied numerically.
\end{abstract}


\maketitle

{\it 1. Introduction.} --- Flows of BECs past obstacles play an important role in the superfluidity physics.
At the initial stage of the development of the superfluidity theory such kind of problems permitted Landau to formulate
his famous superfluidity criterion and two-fluid hydrodynamics model explaining different viscosity
properties of liquid HeII in different experimental setups \cite{landau-41,landau-47}. Physically,
Landau criterion corresponds to the threshold value of the flow velocity for generation of
linear quasiparticles---phonons and rotons in HeII. In reality, the loss of superfluidity can occur
at smaller flow velocity which corresponds to the threshold of generation of nonlinear excitations,
e.g., vortices or vortex rings in HeII \cite{feynman}, and this effect has been intensely studied
theoretically in the model of weakly nonlinear Bose gas and realized experimentally in experiments
with cold atoms (see, e.g., \cite{anglin-01, aftalion-03} and references therein). It was found
that the critical velocity at which superfluid behavior breaks down is about $V_c\cong0.35c_s$,
where $c_s$ denotes the sound velocity in a
uniform condensate far enough from the obstacle. For the flow velocity above the sound velocity, $V>c_s$,
a new channel of dissipation opens---Cherenkov radiation of Bogoliubov waves whose interference
results in formation of a specific ``ship wave'' pattern located, due to the properties of the Bogoliubov dispersion relation,
outside the Mach cone \cite{caruso,gegk,gk-07,ship2}. Emission of vortices dominates in the
interval of velocities $c_s<V<1.43c_s$ where vortices form so-called ``vortex streets''
located inside the Mach cone. For velocities $V>1.43c_s$ the vortex streets transform into oblique
dark solitons \cite{egk1} which become effectively stable with respect to decay into vortices
due to transition from absolute instability of dark solitons to their convective instability in
the reference frame related with the obstacle \cite{kp08,kk-11,hi-12}. Such oblique solitons have
been realized in experiments with polariton condensates \cite{amo-2011,grosso-2011}.

The variety of wave patterns generated by the flow of condensate past an obstacle becomes even richer
in the case of two-component condensates in which two different kinds of motion become possible---density
modes with in-phase motion of the components and polarization modes with their counter-phase motion
(see, e.g., \cite{Fil05,gladush-2009}). Correspondingly, in the linear regime there are two types of
sound waves with different sound velocities $c_d$ and $c_p$ and different Mach cones. Consequently,
two partially overlapping ship wave patterns are generated by the flow \cite{gladush-2009}.
In nonlinear regime, the density and polarization modes demonstrate quite different behavior:
there exist one type of nonlinear density excitations (density solitons) and a number of nonlinear
polarization excitations (solitons, breathers, kinks); see, e.g. \cite{kklp-13}. Not all of them
can be generated by the flow of the two-component BEC past an obstacle with a given type of the potential.
For example, if the obstacle repels both types of species in the two-component BEC, then only the
oblique density solitons are generated and no nonlinear polarization excitations have been
observed  \cite{gladush-2009}. On the contrary, if the
obstacle is polarized, i.e. it repels one component and attracts the other one, then the
oblique breathers are generated instead of the oblique solitons \cite{Kam2013}. This observation
provides potentially a method of distinguishing different types of the obstacle potentials.

In this Communication we demonstrate that the behavior of the wave patterns generated by the flow
past an obstacle can also crucially depend on the properties of the spin-orbit (SO) coupling between 
two components of the condensate. Such a coupling can be induced by creation of artificial gauge
fields in ultracold atomic gases (see, e.g., review article \cite{dgjo-11} and reference therein).
In this method, the two-component $\psi$-function of the condensate corresponds to a spinor field
and the spin-orbit interaction depends on the configuration of atomic levels coupled with laser fields.

{\it 2. The model.} --- We shall consider the 2D-geometry when all variables depend on the
coordinates $\mathbf{r}=(x,y)$ only. The flow is directed along the $x$-axis
and the SO coupling is anisotropic with the Hamiltonian
\begin{equation}\label{eq1}
    H_{SO}=\frac12\left[(p_x-\gamma_x\sigma_y)^2+(p_y-\gamma_y\sigma_x)^2\right]
\end{equation}
written here in standard non-dimensional units; $\sigma_y$ and $\sigma_x$ are Pauli matrices.
In what follows we assume that the SO coupling is tuned in such a way that  $|\gamma_x|\ll|\gamma_y|$
and, hence, the $x$-component of SO coupling can be neglected in the main approximation. As a result,
an incident flow does not ``feel'' this interaction. However, when the flow is disturbed by the obstacle,
then the $y$-component of the flow velocity in the wave motion interacts with the gauge field
that results in drastic modification of the entire wave pattern as we show below.

Taking standard Gross-Pitaevskii (GP) equations for the two-component BEC, we modify them by
account of the SO interaction and obtain the system ($\ga_y\equiv\ga$)
\begin{equation}\label{eq2}
    \begin{split}
    i\prt_t\psi_1+\frac12\Delta\psi_1-(g_{11}|\psi_1|^2&+g_{12}|\psi_2|^2)\psi_1\\
    &-i\gamma\prt_y\psi_2=\kappa_1V_{\mathrm{obs}}(\mathbf{r})\psi_1,\\
    i\prt_t\psi_2+\frac12\Delta\psi_2-(g_{12}|\psi_1|^2&+g_{22}|\psi_2|^2)\psi_1\\
    &-i\gamma\prt_y\psi_1=\kappa_2V_{\mathrm{obs}}(\mathbf{r})\psi_2,
    \end{split}
\end{equation}
where $\kappa_{1,2}V_{\mathrm{obs}}(\mathbf{r})$ denotes the obstacle potential which is repulsive for both components
if $\kappa_1=\kappa_2=1$ and it is repulsive for $\psi_1$-component and attractive for $\psi_2$-component
if $\kappa_1=-\kappa_2=1$. The nonlinear interaction constants are positive and we suppose for simplicity
that $g_{11}=g_{22}\equiv g_1>g_{12}\equiv g_2$.

It is easier to study the excitations with the use of such variables that the corresponding motions are
separated. To this end we introduce a spinor representation of the field variables \cite{ktu-2005}
\begin{equation}\label{eq3}
    \left(
            \begin{array}{c}
              \psi_1 \\
              \psi_2 \\
            \end{array}
          \right)=
          \sqrt{\rho}e^{i\Phi/2}
          \left(
            \begin{array}{c}
              \cos\frac{\theta}2\,e^{-i\phi/2} \\
              \sin\frac{\theta}2\,e^{i\phi/2}  \\
            \end{array}
          \right),
\end{equation}
where $\Phi$ has the meaning of the velocity potential of the in-phase motion ($\mathbf{U}=\nabla\Phi$)
related with variations of the total density $\rho=|\psi_1|^2+|\psi_2|^2$;
the angle $\theta$ is the variable describing the relative density of
the two components ($\cos\theta=(\rho_1-\rho_2)/\rho$, $\rho_1=|\psi_1|^2=\rho\cos^2(\theta/2)$,
$\rho_2=|\psi_2|^2=\rho\sin^2(\theta/2)$)
and $\phi$ is the potential of the relative counter-phase motion ($\mathbf{v}=\nabla\phi$).
In these variables the GP equations (\ref{eq2}) take the form
\begin{equation}\label{eq4}
    \begin{split}
    \rho_t&+\frac12\nabla[\rho\,(\nabla\Phi-\cos\theta\nabla\phi)]-\gamma(\rho\sin\theta\cos\phi)_y=0,\\
    \Phi_t&+\frac{(\nabla\rho)^2}{4\,\rho^2}-\frac{\Delta\rho}{2\,\rho}
    -\frac{\cot\theta}{2\,\rho}\nabla(\rho\nabla\theta)\\
&+\frac14\,((\nabla\Phi)^2+(\nabla\theta)^2+(\nabla\phi)^2)+(g_1+g_2)\rho\\
&-\frac{\gamma}{\rho\sin\theta}\left[\rho\Phi_y\cos\phi-(\rho\cos\theta\sin\phi)_y\right]=0,\\
    \rho\,\theta_t&+\frac12\,[\nabla(\rho \sin\theta \nabla\phi)+\rho \, \nabla\theta\cdot\nabla\Phi ]\\
    &-\gamma\left[(\rho\cos\theta\cos\phi)_y+\Phi_y\sin\phi\right]=0,\\
    \phi_t&-\frac{\nabla(\rho\,\nabla\theta)}{2\,\rho\sin\theta}+\frac12\,\nabla\Phi\cdot\nabla\phi
-(g_1-g_2)\rho\cos\theta\\
&-\frac{\gamma}{\rho\sin\theta}\left[\rho\Phi_y\cos\theta\cos\phi-(\rho\sin\phi)_y\right]=0,
    \end{split}
\end{equation}
where we omitted the terms with the obstacle potential. If this system is solved then the velocities of the
BEC components are equal to
\begin{equation}\label{eq5}
\begin{split}
    \mathbf{v}_1=\frac12\nabla(\Phi-\phi)=\frac12(\mathbf{U}-\mathbf{v}),\\
    \mathbf{v}_2=\frac12\nabla(\Phi+\phi)=\frac12(\mathbf{U}+\mathbf{v}).
    \end{split}
\end{equation}

We suppose that far enough from the obstacle the BEC components have constant densities and the incident flow
is uniform and directed along the $x$-axis ($\mathbf{v}_1=\mathbf{v}_2=\mathbf{V}=(V,0)$). Their overall
density is equal here to $\rho=\rho_0$ and the ratio of the densities of the components is determined by the
angle $\theta_0$ ($\rho_1^0/\rho_2^0=\cot^2(\theta_0/2)$). Then the state of the BEC flow disturbed by the
obstacle is described by the variables
\begin{equation}\label{eq6}
\begin{split}
    &\rho=\rho_0+\rho',\quad \Phi=-[V^2+\rho_0(g_1+g_2)]t+\Phi',\\
    &\theta=\theta_0+\theta',\quad \phi=\phi',
    \end{split}
\end{equation}
where the primed variables correspond to small deviations from the incident flow parameters and describe
the wave generated by the flow past an obstacle.

We are interested in funding the overall transformation of the wave pattern caused by the SO coupling
between the BEC components. It is known that in the situations without the SO coupling the wave patterns
consist of different regions separated by the Mach cone lines
(see \cite{caruso,gegk,gk-07,ship2,egk1,gladush-2009,Kam2013}). Therefore the general features
of the wave patterns are determined by the positions of the Mach cone lines and here we restrict ourselves 
to analytical treatment of these most important wave characteristics. To this end,
we linearize the system (\ref{eq4}) with respect to the small primed variables and neglect dispersion effects
(note that we have to keep the terms with the second order space derivatives of the potentials $\Phi'$
and $\phi'$ since they correspond to the first order derivatives of the physical variables---velocities
$\mathbf{v}_1$ and $\mathbf{v}_2$). 
Looking for the plane wave solution $\rho',\,\Phi',\,\theta',\,\phi'\propto\exp[i(k_xx+k_yy-\om t)]$
of the linearized equations, we obtain in a usual way the dispersion relation
\begin{equation}\label{eq8}
\begin{split}
    \widetilde{\om}^4&-(g_1k^2+2\ga^2k_y^2)\widetilde{\om}^2+g_2\ga\sin\theta_0k^2k_y\widetilde{\om}\\
    &+\frac14(g_1^2-g_2^2)\sin^2\theta_0k^4+\ga^2(g_1k^2-\ga^2k_y^2)k_y^2=0,
    \end{split}
\end{equation}
where $\widetilde{\om}=\om-Vk_x$, $k^2=k_x^2+k_y^2$. If $\ga=0$, then Eq.~(\ref{eq8}) gives well-known
expressions for the density ($c_d$) and polarization ($c_p$) sound velocities which in our notation
take the form
\begin{equation}\label{eq9}
    c_{d,p}^2=\frac12\left[g_1\pm\sqrt{g_1^2\cos^2\theta_0+g_2^2\sin^2\theta_0}\right].
\end{equation}
The positions of the Mach cone lines can be found in the following way. Such a line represents a stationary
plane wave inclined with respect to the $x$-axis by the angle $\chi_M$. Hence, along it we have
$\om=0$ and $k_x=-k\sin\chi_M,\,k_y=k\cos\chi_M$. Substitution of these values of the parameters
into Eq.~(\ref{eq8}) yields the equation for $\chi_M$ whose four roots determine the positions of
the Mach cone lines in the $(x,y)$-plane.

\begin{figure}[ht]
\includegraphics[width=1.0\linewidth]{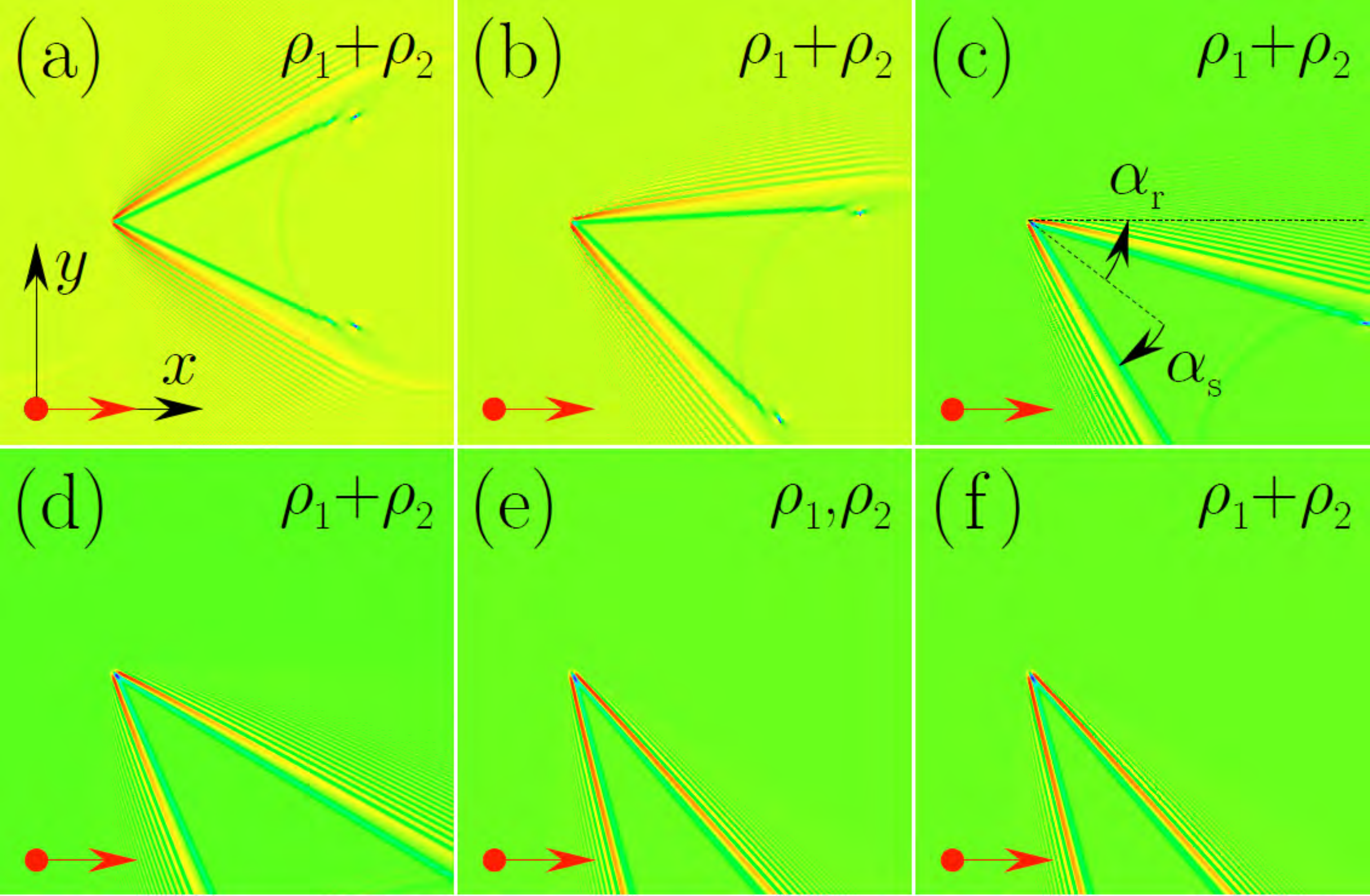}
\caption{(Color online) Transverse density distributions at $t=112$ generated by the flow with velocity
$V=2.6$ past a non-polarized obstacle with $\kappa_1=\kappa_2=1$ at (a) $\ga=0$, (b) $\ga=1$, (c) $\ga=2$,
(d) $\ga=3$, and (e), (f) $\ga=5$. Red arrows indicate the direction of
the flow. Rotation angle $\al_r$ of the entire pattern and the angle between two
oblique dark solitons $\al_s$ are indicated on the panel (c).}
\label{fig1}
\end{figure}

This equation is greatly simplified in the most important case of equal densities of the components,
that is for $\theta_0=\pi/2$. Then Eq.~(\ref{eq8}) easily factorizes and yields the dispersion relations
\begin{equation}\label{eq10}
    \om=Vk_x-\ga k_y\pm c_d^0k,\quad \om=Vk_x+\ga k_y\pm c_p^0k,
\end{equation}
where $c_{d,p}^0$ are given by Eq.~(\ref{eq9}) with $\theta_0=\pi/2$. These relations can be interpreted
as the relations corresponding to the spherical (cylindrical in 2D geometry) waves propagating with
the sound velocities $c_{d,p}^0$ and convected by the flows with effective velocities $(V,\mp\ga)$.
Hence, the first important conclusion is that the density and polarization patterns are rotated due to 
SO coupling in opposite directions by the angles
\begin{equation}\label{eq11}
    \alpha_{r}=\mp\arctan\left(\frac{\ga}V\right),
\end{equation}
where the upper sign corresponds to the density wave and the lower one to the polarization wave.
Besides that, the effective Mach numbers are changed to
\begin{equation}\label{eq12}
    M_{d,p}=\frac{\sqrt{V^2+\ga^2}}{c_{d,p}}
\end{equation}
and the corresponding Mach angle $\al_{c}$ between the Mach lines and their bisectrix is determined now
by the equation
\begin{equation}\label{eq13}
    \sin\al_{c}=\frac1{M_{d,p}},
\end{equation}
and it is different again for the density and polarization waves. The angles $\chi_M$ introduced above 
are therefore given by $\chi_M=\al_r\pm\al_c$.
These expressions completely determine the unusual transformations of the linear wave patterns under the action of the
SO interaction between BEC components. To extend these predictions to the nonlinear waves, we shall resort 
to the numerical solutions of the GP equations (\ref{eq2}).

{\it 3. Numerical simulations.} ---  We have studied
exact numerical solutions of the GP equations (\ref{eq2}) with two-fold aim:  (i) to determine the
obstacle potentials corresponding to generation of the density or polarization waves in SO-coupled BEC; 
(ii) to find out what happens with nonlinear excitations
(oblique solitons and breathers) under the action of SO interaction.

First, we consider the flow past a non-polarized obstacle with $\kappa_1=\kappa_2=1$ and typical results are
shown in Fig.~1 for different values of $\ga$. As one can see, in all cases only density waves are generated
and the entire wave pattern rotates in the direction predicted by Eq.~(\ref{eq11}) with negative
$y$-component of the effective ``flow velocity'' equal to $\ga$. The Mach cone narrows down with growth of $\ga$.
The absence of the polarization mode follows from identity of the wave patterns for different components
shown in Fig.~1(e) and their coincidence with the wave pattern for the overall density shown in Fig.~1(f).

\begin{figure}[ht]
\includegraphics[width=1.0\linewidth]{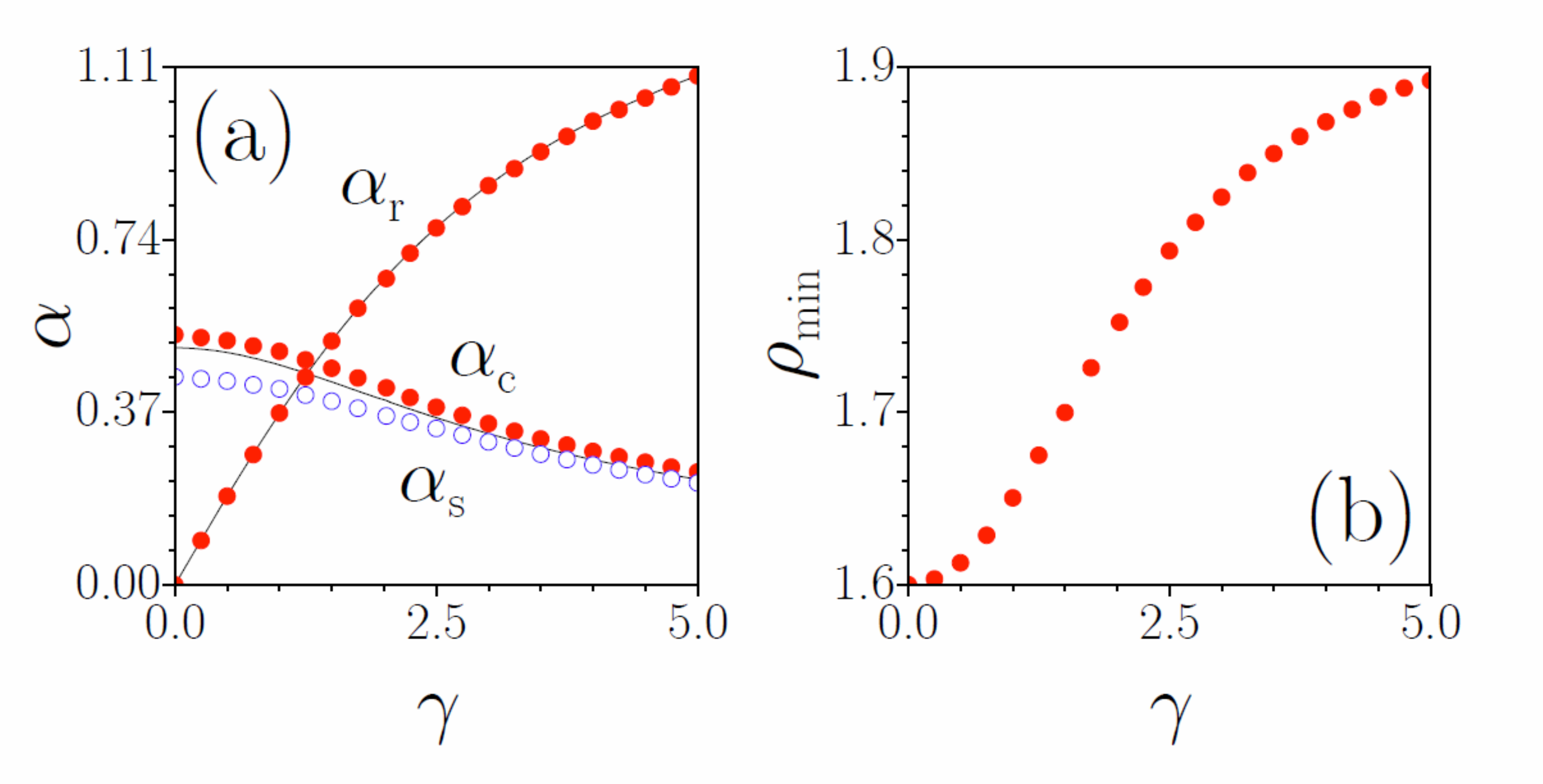}
\caption{(Color online) (a) Mach cone angle $\al_c$ and cone rotation angle $\al_r$ (solid red circles),
as well as half of the angle between oblique dark solitons $\al_s$ (open blue circles) versus spin-orbit
coupling strength $\ga$ in the case of non-polarized obstacle with $\kappa_1=\kappa_2=1$. Solid lines
show analytical prediction, while circles show numerical results. (b) Depth of oblique dark solitons
generated in the flow past non-polarized obstacle versus $\ga$. In all cases $V=2.6$.}
\label{fig2}
\end{figure}

\begin{figure}[ht]
\includegraphics[width=1.0\linewidth]{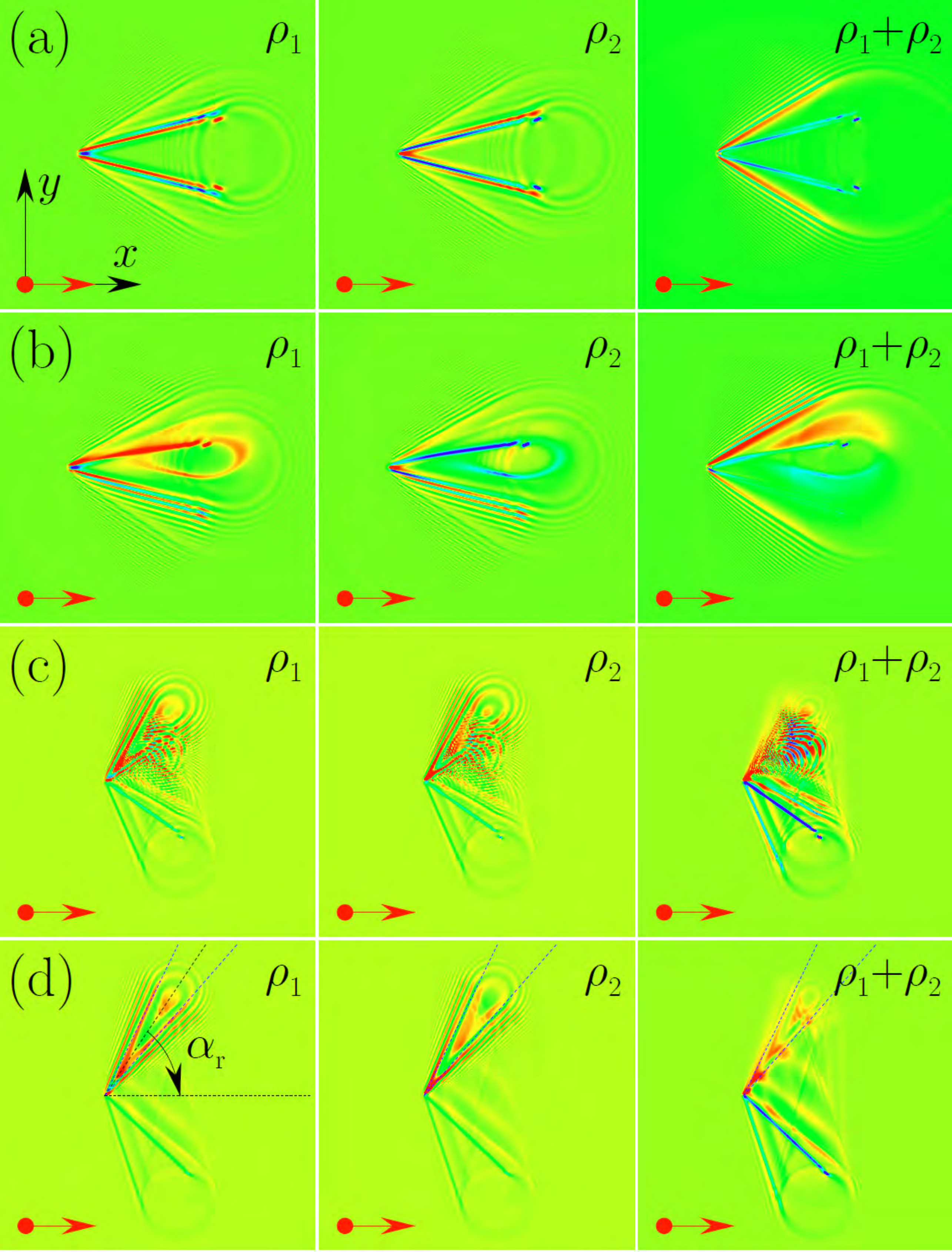}
\caption{(Color online) Transverse density distributions at $t=64$ (a),(b) and $t=32$ (c),(d) generated by the flow with velocity
$V=2.6$ past a polarized obstacle with $\kappa_1=-\kappa_2=1$ at (a) $\ga=0$, (b) $\ga=0.1$, (c) $\ga=3$,
and (d) $\ga=4$. Red arrows indicate the direction of
the flow. The transverse scales in (a),(b) and (c),(d) are different.}
\label{fig3}
\end{figure}

To perform quantitative comparison of numerical results with analytical formulae, we have plotted the
dependence of the rotation angle $\al_r$ (defined geometrically in Fig.~1(c)) on $\ga$ in Fig.~2(a).
The numerical dependence perfectly agrees with the analytical formula (\ref{eq11}). The position
of the Mach cone cannot be defined unambiguously in numerical plots because the analytical formulas
refer to the dispersionless limit of very long wavelength but in reality the wavelength of typical
excitations is finite and the transition from the region ``inside'' the Mach cone to the region
``outside'' it is quite smooth.  Therefore we use a practical definition of the position of the
Mach cone line as a line of maxima of the total density close to the region of the ship wave
oscillations. The corresponding numerical values of $\al_c$ are shown in Fig.~2(a) and they
slightly exceed analytically predicted Mach cone angle (\ref{eq13}). Besides that,
we have determined the angle $\al_s$ between the oblique soliton and the bisectrix of the Mach cone
(see Fig.~1(c)). Its dependence on $\ga$ is also shown in Fig.~2(a). As one can see the angle $\alpha_s$ 
approaches the analytically predicted Mach cone angle with increase of $\gamma$. This means that
the oblique solitons become more shallow with increase of $\ga$ and such a behavior of the
soliton's depth is corroborated by numerical data plotted in Fig.~2(b). According to the
analytical theory (see, e.g., \cite{gladush-2009}) the soliton depth is a function
of its squared total velocity, that is, in our case, of the combination $V^2+\ga^2$.
Consequently, for small $\ga$, $|\ga|\ll V$, we get parabolic dependence of the soliton depth
on $\ga$ in agreement with plot in Fig.~2(b). Our numerical results show that oblique
solitons are quite robust with respect to the influence of the SO interaction.

At last, we have studied the dependence of the wave patterns generated by the flow past a
polarized obstacle on the SO coupling constant. The results are shown in Fig.~3. Although both
density and polarization ship waves are generated, only the polarization oblique breather
is formed now by the polarized obstacle. Since the density
and polarization wave patterns have different structures and they are rotated by the SO
interaction in opposite directions, the entire wave pattern becomes asymmetric. For small
$\ga$ these two types of waves overlap that leads to considerable deformations of
oblique breahters (see Fig.~3(b)). For large $\ga$ the density and polarization waves
are separated due to large values of the rotation angles $\al_r$ that agree very
well with the analytical formulae (\ref{eq11}). However, the oblique breathers demonstrate
considerable sensitivity to disturbances created by the interference with density waves,
especially at large evolution times, 
[Fig.~3(c)], and only for larger values of $\ga$ of the order of $\ga\approx4$ the polarization 
and density cones are very well resolvable in the wave pattern [Fig.~3(d)]. 
Thus, the oblique breathers are more fragile structures than the oblique solitons.

Summarizing, we predicted that density and polarizations wave structures
generated by the flows of two-component condensates past obstacles are rotated by
the SO coupling (\ref{eq1}) in opposite directions. The considerable narrowing of the Mach cones 
upon increase of SO coupling strength is also demonstrated.

\end{document}